\documentclass[epjCONF]{svjour}
\usepackage{graphics}
\usepackage[varg]{txfonts} 
\usepackage[latin1]{inputenc}
\session-title{12th International Workshop on Meson Production, Properties and Interaction}
\begin{document}
\title{Dalitz plot studies of \mbox{\boldmath $D^0 \to K^0_S \pi^+ \pi^-$} decays}
\author{L. Le\'sniak\inst{1}\fnmsep\thanks{\email{Leonard.Lesniak@ifj.edu.pl}} 
\and R. Kami\'nski\inst{1}
\and A. Furman\inst{2}
\and J.-P. Dedonder\inst{3}
\and B. Loiseau\inst{3}
}
\institute{Division of Theoretical Physics, The Henryk Niewodnicza\'nski Institute of Nuclear Physics,
                  Polish Academy of Sciences, 31-342 Krak\'ow, Poland
 \and ul. Bronowicka 85/26, 30-091 Krak\'ow, Poland
 \and Laboratoire de Physique Nucl\'eaire et de Hautes \'Energies, Groupe Th\'eorie, 
Universit\'e Pierre et Marie Curie et Universit\'e Paris-Diderot, IN2P3 et CNRS, 4 place Jussieu,
 75252 Paris, France}
\abstract{ 
The availaible data on the $D^0 \to K^0_S \pi^+ \pi^-$ decays are
analysed in the framework of the quasi two-body QCD factorization 
approximation. The annihilation, via W-exchange, amplitudes are added to the weak-decay
 tree amplitudes. The doubly Cabibbo suppressed parts of the amplitudes are also
considered. The strong interactions between the kaon-pion and pion-pion pairs in 
the $S$-, $P$- and $D$-final states are described in terms of the 
corresponding form factors. 
The kaon-pion or pion-pion scalar and vector form factors are 
constrained by other experimental data.
Unitarity, analyticity and chiral symmetry are also used to obtain
 their functional forms.
We go through a minimization 
procedure to reproduce the  $K^0_S \pi^-$, $K^0_S \pi^+$ and 
$\pi^+ \pi^-$ effective mass projections of the Dalitz plot distributions.
The large number (27) of non-zero amplitudes leads to a large number of parameters.
The resulting model distributions and branching fractions are 
compared to the accurate Belle Collaboration data.
}
%
\maketitle
\section{Introduction}
\label{intro}
Dalitz-plot time dependent amplitude analyses have been recently  
performed \cite{BaBar,Belle} for the $CP$ 
self conjugate $D^0$ meson decays into $K^0_S \pi^+ \pi^-$.
These studies have allowed 
a direct measure of the $D^0$-$\bar D^0$ mixing parameters, 
the knowledge of which could show the presence of new physics 
contribution beyond the standard model.
Studies of 
$B^{\pm}\rightarrow D^{(*)} K^{\pm}, D\rightarrow K^0_S \pi^+ \pi^-$
decays,
in which the interference between $D^0$ and $\bar D^0$ mesons was
used to measure the Cabibbo-Kobayashi-Maskawa angle $\gamma$,
have also been accomplished \cite{BaBar2,Belle2}.
A good understanding of the final state interactions between
mesons in the $D^0$ decays into $K^0_S \pi^+ \pi^-$ is essential 
in order to reduce errors of the $D^0-\bar D^0$ mixing parameters and
the measured values of the  angle $\gamma$. 
A very rich resonance spectrum seen in the Dalitz plots is a direct signal
of the complexity of the strong meson interactions.
Using these high statistics data theoretical models of the decay 
amplitudes can be tested.

The experimental analyses \cite{BaBar,Belle} relied mainly on 
application of an isobar model. In such a model one can accomodate many
resonances coupled to interacting pairs of mesons. However, one should stress that
the corresponding decay amplitudes are not unitary: unitarity is not
preserved in the three-body decay channels and it is also violated in
the two-body subchannels.
Within the isobar model it is particularly difficult to distinguish 
the $S$-wave amplitudes from the non-resonant background terms. 
Their interference is often very strong which means that some two-body branching
fractions, extracted from the data, could be unreliable.
The isobar model is flexible but it has many free parameters (at least two fitted parameters
for each amplitude component). For example, the Belle Collaboration has used 40 fitted
parameters in ref.~\cite{Belle} and the Babar Collaboration 43 free parameters in ref.~\cite{BaBar}.

The construction of unitary three-body strong interaction amplitudes in a wide range of
meson-meson effective masses is difficult. Therefore, as a first step, we incorporate 
two-body unitarity in our model of the $D$-decay amplitudes with final state interactions 
in the 
 $K^0_S \pi^{\pm}$ $S$-  and $P$-waves 
and in the $\pi^+\pi^-$ $S$-wave.
The branching fraction corresponding to a sum of these amplitudes is close to 80 \%
of the total branching fraction of the $D\rightarrow K^0_S \pi^+ \pi^-$ decay.

\section{Decay amplitudes}
\label{sec:1}
The $D^0$ decays into $K^0_S \pi^+ \pi^-$ are analysed in the framework of the quasi-two
body factorization approach.
The decay amplitude consists of the following 27 non-zero   parts:
seven \textit{allowed} tree amplitudes, six doubly Cabibbo \textit{suppressed} tree amplitudes,
fourteen \textit{annihilation} (W-exchange) amplitudes (7 allowed and 7 Cabibbo 
suppressed).
The allowed amplitudes are generated by the quark $c \rightarrow s u \bar d$ transitions and
the doubly Cabibbo suppressed amplitudes correspond to  $c \rightarrow d u \bar s$ decays.
Seven partial wave amplitudes are considered. We count $S$, $P$ and $D$ waves
in the $K^0_S \pi^{\pm}$ and the $\pi^+ \pi^-$ subsystems and separately the $P$-wave
amplitude for the $G$- parity violating $\omega \rightarrow \pi^+ \pi^-$ transition.
There are numerous resonances contributing to different partial wave amplitudes.
For example, in the $K^0_S \pi^-$ $S$-wave subchannel one can list $K_0^*(800)^-$ or $\kappa^-$
and $K_0^*(1430)^-$ resonances, in the $P$-wave one $K^*(892)^-$, $K_1(1410)^-$ and $K^*(1680)^-$
and in the $D$-wave the $K_2^*(1430)^-$ resonance. 
The same resonances but 
with the opposite charge are active in the $K^0_S \pi^+$ subchannels. 
In the $\pi^+ \pi^-$ 
subchannels one can enumerate the $f_0(500)$ or $\sigma$, $f_0(980)$ and  $f_0(1400)$ scalars,
the $\rho(770)$, $\rho(1450)$ and $\omega(782)$ vector states and the $f_2(1270)$ tensor
resonance.

The main part of the weak decay effective Hamiltonian is proportional to the following operator:
\begin{equation}
\label{op}
O=\frac{G_F}{\sqrt2}V^*_{cs} V_{ud} j_1 j_2, 
\end{equation}
where $G_F$ is the Fermi coupling constant, $V_{cs}$ and $V_{ud}$ are the 
Cabibbo-Kobayashi-Maskawa matrix elements, the two quark currents are defined as
$j_1=(\bar s c)_{V-A}$ and $j_2=(\bar u d)_{V-A}$,
the symbol $V-A$ means that the axial current is subtracted from the vector current.
In the quasi-two body factorization approximation one can write:
\begin{equation}
\label{fact}
\langle K^0_S \pi^-\pi^+|j_1 j_2|D^0 \rangle \cong 
\langle K^0_S \pi^-|j_1|D^0 \rangle  \langle \pi^+|j_2|0 \rangle
+\langle \pi^- \pi^+|j'_1|D^0 \rangle \langle K^0_S|j'_2|0 \rangle
+\langle 0|j'_1|D^0 \rangle \langle K^0_S \pi^-\pi^+|j'_2|0 \rangle,
\end{equation}
where $|0 \rangle$ denotes the vacuum state and $j'_1=(\bar u c)_{V-A}$, 
$j'_2=(\bar s d)_{V-A}$ are the quark currents obtained from the $j_1, j_2$ currents by
the Fierz transformation. 
This form of the transition matrix element enables one to
introduce the pion $f_{\pi}$, kaon $f_K$ and $D^0$ $f_D$ decay constants. 
The corresponding expressions are:
\begin{equation}
\label{fff}
\langle \pi^+|j_2|0 \rangle =if_{\pi} p_{\pi},
~~~~~~\langle K^0_S|j'_2|0 \rangle =if_K p_K,
~~~~~~\langle 0|j_1'|D^0 \rangle =-if_D p_D,
\end{equation}
where $p_{\pi}$, $p_K$ and $p_D$ are the pion, kaon and $D$ meson four-momenta, respectively.

\subsection{Transition matrix elements}
\label{trans}
The expression of the transition matrix elements can be simplified if we assume that
the two strongly interacting mesons $h_2$ and $h_3$ of momenta $p_2$ and $p_3$ form
a resonance $R$ in the final state.
Then one can write:
\begin{equation}
\label{hhD}
\langle h_2(p_2)h_3(p_3)|j~|D^0(p_D)\rangle = G_{Rh_2h_3}(s_{23})\langle R(p_2+p_3)|j~|D^0(p_D)\rangle,
\end{equation}
where $G_{Rh_2h_3}(s_{23})$ with $s_{23}=(p_2+p_3)^2$ is the vertex function describing the $R$ decay
into $h_2$ and $h_3$ mesons.
For example, let us consider the reaction $D^0(p_D) \rightarrow \pi^+(p_1) K^0_S(p_2) \pi^-(p_3)$.
If in the intermediate state the $K^*(892)^-$ resonance is formed and decays into
a $\bar K^0\pi^-$ pair then the transition form factor $A_0^{DK^{*-}}(m_{\pi}^2)$ appears in the
matrix element
\begin{equation}
\label{RjD}
\langle R(p_2+p_3)|j~|D^0(p_D)\rangle=-2im_{K^*}\frac{\epsilon^* \cdot p_D}{p_1^2}~p_1
~A_0^{DK^{*-}}(m_{\pi}^2) ~~ +3~~other~~ terms.
\end{equation}
Here $j=(\bar s c)_{V-A}$, $\epsilon$ is the $K^*(892)^-$ polarization vector, $m_{K^*}$ is the
$K^*$ mass and $p_D=p_1+p_2+p_3$.          
The ``$3~~ other~~ terms$'' do  not contribute to the transition amplitude in eq. (\ref{fact}).
The vertex function $G_{K^{*-}\bar K^0\pi^-}$ can be expressed in terms of the kaon-pion transition
vector form factor $F_1^{\bar K^0\pi^-}(s_{23})$
\begin{equation}
\label{GK*}
G_{K^{*-}\bar K^0\pi^-}(s_{23})=\epsilon \cdot(p_2-p_3)\frac{1}{m_{K^*}f_{K^*}}F_1^{\bar K^0\pi^-}
(s_{23}).
\end{equation}
In the above equation $f_{K^*}$ is the $K^*$ decay constant.
Assuming isospin symmetry one can equate the transition vector form
factor $F_1^{\bar K^0\pi^-}(s_{23})$ to the charged kaon to charged pion transition vector form
factor $F_1^{K^-\pi^+}(s_{23})$ calculated in ref.~\cite{PR09}.

The third term of eq. (\ref{fact}) can also be simplified if the two hadrons, for example $h_2$ and
$h_3$, interact via a resonant state $R$. Then, similarly to eq. (\ref{RjD}) we write
\begin{equation}
\label{hhhj0}
\langle h_1(p_1)h_2(p_2)h_3(p_3)|j'|0\rangle=G_{Rh_2h_3}(s_{23})\langle h_1(p_1)R(p_2+p_3)|j'|0\rangle.
\end{equation}
If, for example, $h_1=\bar K^0$, $R=f_0\rightarrow \pi^+\pi^-$ and $j'=(\bar s d)_{V-A}$ then the matrix
element reads
\begin{equation}
\label{Kf0}
\langle \bar K^0(p_1)f_0(p_2+p_3)|j'|0\rangle=-i\frac{m^2_{K^0}-s_{23}}{p_D^2}p_D 
F_0^{\bar K^0f_0}(m_D^2)~~~~ +~~~~second~~~ term,
\end{equation}
where $F_0^{\bar K^0f_0}(m_D^2)$ is the kaon to $f_0$ scalar transition form factor.
The vertex function for the $f_0$ decay into $\pi^+\pi^-$ can be parametrized as
\begin{equation}
\label{f0pipi}
G_{f_0\pi^+\pi^-}(s_{23})=\chi_2 F^{\pi+\pi-}_0(s_{23}),
\end{equation}
where $\chi_2$ is a constant and $F^{\pi+\pi-}_0(s_{23})$ is the pion scalar form factor.
Its functional form is taken from our $B^{\pm} \rightarrow \pi^{\pm}  \pi^{\mp}\pi^{\pm}$ decay
study~\cite{APP11}. It preserves unitarity and groups together
 three scalar-isoscalar resonances $f_0(500)$, $f_0(980)$ and $f_0(1400)$. 

\subsection{Examples of allowed transition amplitudes with $K^0_S\pi^-$ final state interactions}
Using the assumptions introduced in subsection~\ref{trans} one can derive formulae for the decay
amplitudes in which the $K^0_S\pi^-$ final state interactions are explicitely included. The 
$S$-wave amplitude reads:
\begin{equation}
\label{S}
A_S=-\frac{G_F}{2}\Lambda_1 a_1 f_{\pi} (m_D^2-m^2_{\pi})~ F_0^{D K_0^{*-}}(m^2_{\pi}) 
~F_0^{\bar K^0\pi^-}(m^2_-).
\end{equation}
Here, $\Lambda_1=V_{cs}^*V_{ud}$, $a_1$ is the effective Wilson coefficient, 
$F_0^{D K_0^{*-}}(m^2_{\pi})$ is the scalar 
$D^0$ to $K_0^{*-}$ transition form factor
and $F_0^{\bar K^0\pi^-}(m_-^2)$ is the scalar $\bar K^0$ to $\pi^-$ transition form factor
which depends on the $K^0_S\pi^-$ effective mass squared $m^2_-$.
The latter form factor can be taken equal to the $K^-$ to $\pi^+$ transition scalar form factor 
$F_0^{K^-\pi^+}(m^2_{\pi})$ which has been evaluated in the study of 
$B \rightarrow K \pi^+\pi^-$ decays ~\cite{PR09}. 
In this form factor 
both $K^*_0(800)$ and $K^*_0(1430)$ resonances are included in a unitary way.
The expression for the $P$-wave amplitude is:
\begin{equation}
\label{P}
A_P=-\frac{G_F}{2}\Lambda_1 a_1 \frac{f_{\pi}}{f_{K^*}}[m_0^2-m_+^2+\frac{(m_D^2-m_{\pi}^2) 
(m_K^2-m_{\pi}^2)}{m_-^2}]~A_0^{DK^{*-}}(m_{\pi}^2)~F_1^{\bar K^0\pi^-}(m^2_-),
\end{equation}
where $m^2_+$ is the $K^0_S\pi^+$
effective mass squared and $m^2_0$ denotes the $\pi^+ \pi^-$
effective mass squared.
The $D$-wave decay amplitude depends on the mass $m_{K_2^*}$ and the width $\Gamma_{K_2^*}$
of the resonance $K_2^*(1430)$:
\begin{equation}
\label{D}
A_D = -\frac{G_F}{2}\Lambda_1 a_1 f_{\pi} F^{DK_2^*}(m^2_-)\frac{g_{K_2^*K_S^0\pi}
D(m_+^2,m_-^2)}{m^2_{K_2^*}-m_-^2-im_{K_2^*}\Gamma_{K_2^*}}.
\end{equation}
Here, $F^{DK_2^*}(m^2_-)$ is a combination of three types of the $D$ to $K_2^*(1430)^-$
transition form factors, $g_{K_2^*K_S^0\pi}$ is the decay coupling constant and the $ D(m_+^2,m_-^2)$
is the $D$-wave angular distribution function.

 Derivation of other decay amplitudes proceeds in a quite similar way as shown above for the 
$A_S$, $A_P$ and $A_D$ amplitudes. One can notice that in our amplitude model the meson transition 
form factors play a very important role.
\section{Results}
\label{res}
The theoretical model outlined in the previous section has 28 free real parameters, most of them are
unknown complex values of the meson-meson transition form factors appearing in the fourteen W-~exchange 
amplitudes. 
The scalar and vector kaon-pion and scalar pion form factors are constrained using 
unitarity, analyticity and chiral symmetry. Fig. 1 shows that the model preliminary results are in 
fair agreement with the Belle Collaboration data.
Also the total braching fraction is well reproduced ~\cite{PDG}.

\begin{figure}
\begin{center}
\resizebox{0.52\columnwidth}{!}{
\includegraphics{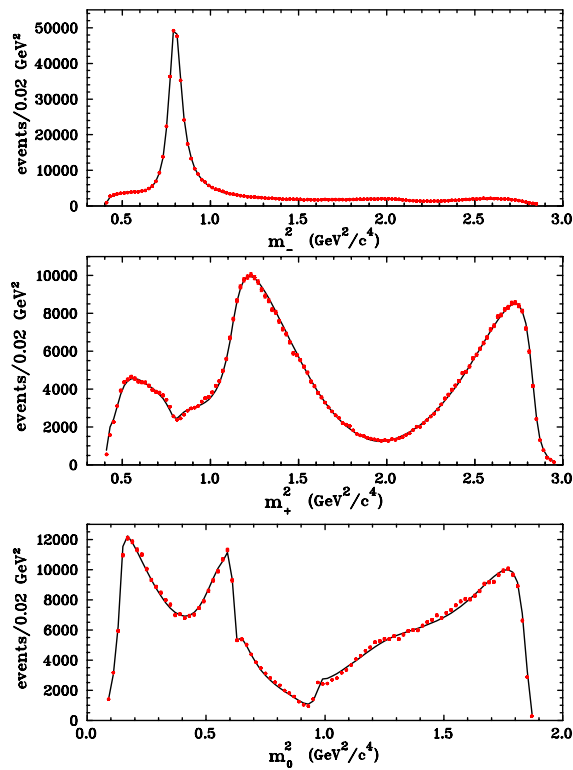} }
\end{center}
\label{rys}
\caption{Comparison of the Dalitz-plot projections for the Belle data ~\cite{Belle}
with the present model.}
\end{figure}
%
\section{Acknowledgements}
This work has been supported in part by the IN2P3-Polish Laboratories Convention
(project No 08-127).


\begin{thebibliography}{}
\bibitem{BaBar}
P. del Amo Sanchez \textit{et al.} (BABAR Collaboration),
Phys. Rev. Lett. \textbf{105}, (2010) 081803  \\ and arXiv: hep-ex 1004.5053v3
\bibitem{Belle} 
L. M. Zhang,  \textit{et al.} (Belle Collaboration),
Phys. Rev. Lett. \textbf{99}, (2007) 131803
\bibitem{BaBar2} 
P. del Amo Sanchez \textit{et al.} (BABAR Collaboration),
Phys. Rev. Lett. \textbf{105}, (2010) 121801
\bibitem{Belle2}
A. Poluektov \textit{et al.} (Belle Collaboration),
Phys. Rev. \textbf{D 81}, (2010) 112002
\bibitem{PR09}
B. El-Bennich \textit{et al.}, Phys. Rev. \textbf{D 79}, (2009) 094005
\bibitem{APP11}
J.-P. Dedonder \textit{et al.}, Acta Physica Polonica \textbf{42}, (2011) 2013
\bibitem{PDG}
K. Nakamura \textit{et al.} (Particle Data Group), Journal of Physics G \textbf{37},
(2010) 075021

\end{thebibliography}
\end{document}